\documentstyle[12pt]{article}
\newcommand{\be}{\begin{equation}}
\newcommand{\ee}{\end{equation}}
\newcommand{\br}{\begin{eqnarray}}
\newcommand{\er}{\end{eqnarray}}
\newcommand{\bq}{\begin{eqnarray*}}
\newcommand{\eq}{\end{eqnarray*}}
\newcommand{\lt}{\left}
\newcommand{\rt}{\right}
\newcommand{\td}{\tilde}
\newcommand{\th}{\theta}
\newcommand{\gm}{\gamma}
\newcommand{\sg}{\sigma}
\newcommand{\al}{\alpha}
\newcommand{\pr}{\prime}
\newcommand{\no}{\nonumber}
\newcommand{\pl}{\partial}
\newcommand{\bm}{\bibitem}
\begin{document}
\rightline{SINP/TNP/94-16}
\bigskip
\begin{center}
{\Large{\bf BARYOGENESIS AND DAMPING IN NONMINIMAL ELECTROWEAK MODELS}}\\
\vspace{1cm}
A. Kundu and S. Mallik\\
\smallskip
Saha Institute of Nuclear Physics\\
1/AF, Bidhannagar, Calcutta 700064, India
\end{center}
\vspace{1cm}
\begin{abstract}
We study the effect of damping on the generation of baryon asymmetry of
the Universe in the standard model of the eletroweak theory with simple
extensions of the Higgs sector. The propagation of quarks of masses up to
about 5 GeV are considered, taking into account their markedly different
dispersion relations due to interaction with the hot electroweak plasma.
It is argued that the contribution of the b quark can be comparable to
that of the t quark calculated earlier.
\end{abstract}
\vspace{1cm}
{\small {PACS number(s): 98.80 Cq, 12.15 Ji}}
\newpage
\section{Introduction}
\smallskip
The discovery of baryon number violation in the standard electroweak theory
\cite{'thooft} has led to the possibility of constructing a scenario for
baryogenesis at the electroweak scale in the early universe. This violation,
although exceedingly small at the present epoch, can be unsuppressed at the
then prevailing high temperature \cite{krs}. The other two Sakharov
conditions \cite{sakh}
could also be met: C and CP violation originate from the interaction of
quarks with the Higgs fields. Also, if the electroweak phase transition
is of first order, the motion of the walls of bubbles of broken phase
within the unbroken medium would produce the required departure from
thermal equlibrium.

A natural and generic mechanism for electroweak baryogenesis was proposed
by Cohen, Kaplan and Nelson (CKN) \cite{ckn',ckn}. Here a CP-odd charge (like
the
lepton number) is separated by the reflection and transmission of fermions
by the bubble wall. It is then converted into an asymmetry in baryon number
by the baryon number violating processes occuring outside the bubble. As a
straightforward calculation of the observed baryon asymmetry requires too
big a CP violation  to be available in the minimal standard model (MSM),
these authors chose to work with simple extensions of the minimal version.
The required CP violation is achieved by the complex space dependent
fermionic mass function within the bubble wall, which arise quite generally
in such models \cite{ghkd}.

Shaposhnikov \cite{shp} studied a similar mechanism of direct separation of
baryonic number by the bubble wall, taking into account the effect of the hot
ambient plasma on the quark propagation. As a result of interaction with the
quanta of the medium, the quarks acquire temperature dependent effective
masses and satisfy altered dispersion relations. The corresponding fermionic
modes or quasiparticles have very different reflection and transmission
coefficients in different regions of momenta. The details are worked out in
Farrar and Shaposhnikov (FS) \cite{fs}. It is found that quark momenta
relevant for baryogenesis are much lower than the temperature of phase
transition. As a consequence, the CP violation of the MSM model suffices to
generate the observed baryon asymmetry of the Universe.

Recently Gavela {\it et al} \cite{ghop} and Huet and Sather \cite{hs}
object to the FS analysis, pointing out that they do not include the imaginary
part of the effective quark mass giving rise to damping of amplitudes.
Including the damping they find the reflection coefficients and hence the
baryon asymmetry to reduce to a negligible value.

This objection is not agreed upon by FS \cite{fsn}, however. According to them,
although the leading quark interaction with the medium can be simulated by
the real effective mass with an alterted dispersion relation, higher order
interaction giving rise to the imaginary part in the mass and hence damping
cannot be treated in this way, as it gives non-unitary description. Instead,
such dissipative processes must be treated in a many body context.

Pending a proper discussion of the many body problem, it is useful to study
the dissipative quantum mechanical problem itself in connection with the
baryon asymmetry. Here we consider simple extensions of MSM following the
work of CKN. However, while they considered the t quark propagation, we
consider the propagation of lighter quarks, for which the finite temperature
corrections are important. The reflection and transmission amplitudes are
obtained in an iterative series in powers of the mass parameters. The results
for reflection and transmission currents confirm the loss of unitarity for
finite wall thickness. However, as the wall thickness tends to zero, unitarity
is restored, but the fermion-antifermion asymmetry is lost in the model. We
reintroduce the asymmetry by assuming a very large imaginary part of the mass
function within the wall and carry out a representative calculation of the
baryon asymmetry.

In sec. 2 we review the propagation properties of quark excitations in the
electroweak plasma taking damping into account. In sec. 3 we solve the Dirac
equation within the bubble wall where the mass function is space dependent.
It provides the matching condition to be used in sec. 4 to find the
reflection and transmission amplitudes.
In sec. 5 we calculate the baryon asymmetry in the thin wall approximation but
retaining the imaginary part of the mass function. Finally in sec. 6 we
conclude with a discussion of the results obtained.

\vspace{.8cm}
\section{Quark propagation in hot plasma}
\smallskip
In this section we collect the properties of light quark excitations as they
propagate through the electroweak plasma at about the phase transition
temperature. The most important effect of the medium on the quark is obtained
by calculating the quark propagator at finite temperature
\cite{fs,wel}.
There arises a temperature dependent, chirally invariant complex mass with a
modified dispersion relation. Neglecting the electroweak contributions
compared to that due to strong interaction, the leading contribution to the
real part $E_0$ and imaginary part $\gm$ of the effective mass are the same
for both the left($L$)- and right($R$)-handed quasiparticles,
\[E_0=(2\pi\al_s/3)^{1/2}T \simeq .5T \]
and \[\gm=.15\al_sT \simeq .2T\]
with $\alpha_s=.12$ at the Z boson mass. For excitations close to $E_0$, the
effective Lagrangian incorporating the altered dispersion relation is
\cite{fs,ghop,hs}
\be
{\cal L}=iR^{\dagger}(\pl_0+{1\over 3}\sigma\cdot\nabla+iE_0+\gm)R
+iL^{\dagger}(\pl_0-{1\over 3}\sigma\cdot\nabla+iE_0+\gm)+
mL^{\dagger}R+m^{\ast}R^{\dagger}L,
\ee
where we have also included the quark mass acquired through Higgs mechanism.

Having incorporated the effects of field degrees of freedom in (1), the problem
reduces to quantum mechanics of left- and right-handed quasiparticles, having
the structure of a resonance of width $\gm$. We prefer to make the momentum
variable complex rather than the energy, so that the spatial propagation will
be damped.

In the following we consider the one dimensional problem where the
quasiparticles
propagate along the $z$-axis, normal to the bubble wall. Writing
\[L=\lt(\psi_1\atop\psi_2\rt),\qquad R=\lt(\psi_3\atop\psi_4\rt)\]
the equations of motion derived from the Lagrangian (1) split into two
independent sets. Defining
\[\Phi=\lt(\psi_1\atop\psi_3\rt),\qquad \Phi^{\prime}=
\lt(\psi_4\atop\psi_2\rt),\]
and considering solutions of positive energy $E$, they are
\be {d\Phi\over dz}=iQ(z)\Phi \ee where \be Q(z)=3\lt({E-E_0+i\gm\atop
-m(z)}\qquad
{m^{\ast}(z)\atop -(E-E_0+i\gm)}\rt) \ee
and a similar one for $\Phi^{\prime}$ with $m$ replaced by its complex
conjugate. It suffices for us to work with $\Phi$ only. Note that the current
along the $z$-axis carrried by the components $\psi_1$ and $\psi_3$ of $\Phi$
is \be j_z=\Phi^{\dagger}\sg_3\Phi,  \ee $\sg_3$ being the third Pauli matrix.

The planar bubble wall has a finite thickness, extending from $z=0$ to
$z=z_0$ \cite{fn1}.
It separates the broken phase $(z>z_0)$ from the unbroken phase $(z<0)$. The
real part of the Higgs induced mass $m(z)$ rises from zero in the unbroken
phase through the bubble wall to the (almost) zero temperature mass $m_0$ in
the broken phase. The imaginary part is non-zero only within the bubble wall.
Their actual shapes will be conveniently chosen later in sec. 3 below.

In the unbroken phase $(m=0)$, the components $\psi_1$ and $\psi_3$ (also
$\psi_2$ and $\psi_4$) decouple. Consider (damped) plane waves along $z$
direction,$\psi_{1,3}\sim e^{iKz},\qquad K=k+i\Gamma_u,k>0$. They satisfy the
dispersion relations \cite{fn2},
\be E_\pm=E_0\pm {k\over 3}, \Gamma_u=\pm 3\gm , \ee
the (+) and(-) relations holding for $\psi_1$ and $\psi_3$ respectively. In
contrast to the situation for a free massless particle at zero temperature
satisfying $E_\pm=\pm k$, here a part of the (-) branch $(0<E_-<E_0)$ in (5)
is also available for quasiparticle propagation with positive energy.
The ($\pm$) branches are called the normal and abnormal ones respectively.
Unlike the energy $E_\pm$, the variable $k$, however, does not represent the
true momentum ${\bar k_\pm}$ of the excitation, the latter being given by
\be {\bar k_\pm}=\pm {1\over 3}E_\pm  \ee
The reversal of sign of ${\bar k}_-$ in the abnormal branch is in conformity
with the same for the group velocity,
\be v_\pm ={\pl E_\pm\over \pl k}=\pm{1\over 3} \ee
Thus the above solution for $\psi_1$ $(\psi_3)$, belonging to the normal
(abnormal) branch, advances in the positive (negative) $z$-direction. For
solutions $\psi_{1,3}\sim e^{-iKz}$, $\psi_1$ and $\psi_3$ belong to opposite
branches. Note that the propagation is always damped.

In the broken phase, the propagation properties are similar except that the
Higgs $m_0$ couples the two components in $\Phi$ (and in $\Phi^{\prime}$).
Again consider damped plane wave along $z$-direction, $\Phi\sim\chi e^{iPz},
P=p+i\Gamma_b,p>0$. The spinor $\chi$ satisfies
\be \lt({3(E-E_0+i\gm)-P\atop -3m_0}\qquad {3m_0\atop -3(E-E_0+i\gm)-P}
\rt)\chi =0 \ee
We get the dispersion relations by setting the determinant of the $2\times 2$
matrix equal to zero. Separating the real and imaginary parts, we get
\be (E-E_0)^2-(p/3)^2={m_0(p/3)^2\over \gm^2+(p/3)^2},
\qquad \Gamma_b={3\gm(E-E_0)\over p}.\ee
The normal (+) and abnormal (-) branches arise on taking the square root,
\[ E=E_0\pm {p\over 3}g(p),\qquad \Gamma=\pm\gm g(p),\] where
$g(p)=\sqrt{1+{m_0^2 \over \gm^2+(p/3)^2}}.$
Note that the presence of damping $(\gm\neq 0)$ removes any gap between
the normal and the abnormal branches, which exists for $\gm=0$, the
dispersion relation then reducing to $E=E_0\pm\sqrt{p^2/9+m_0^2}$.
The spinor $\chi$ is obtained by solving (8). Normalizing to unit current,
we get \be \chi=\lt(c\atop -s\rt),\ee
where the components $c$ and $s$ stand for
\be c=\cosh \th.e^{i\phi},\; s=\sinh \th.e^{i\phi} \ee
where \[\cosh \th=\sqrt{{E-E_0+p/3\over p/3}},\qquad
e^{4i\phi}={p/3+i\gm\over p/3-i\gm}\]
The presence of non-zero damping also brings in the phase $\phi$.

We note here for later use the Lorentz invariant expression for the density
of fermionic excitations,
\[ n=(exp\: \beta p\cdot v+1)^{-1}\] where $\beta$ is the inverse
temperature of the fluid in the frame where it is at rest, $p^\mu$ is the
energy-momentum 4-vector of the excitation and $v^\mu$, the 4-velocity of
the medium. In the wall rest frame, $p^\mu=(E,\bar{p}), v^\mu=\gm(1,v)$
where $\gm=1/\sqrt{1-v^2}$ and $\bar{p}$ is the true momentum also given by
(6).
For $p$ along the positive $z$-direction, we thus have in this frame,
$p\cdot v=E_\pm(1\mp v/3)$, up to linear term in $v$. In the following we
require the densities of particles with true momenta towards the wall. In the
unbroken phase these are given by
\be n_\pm^u={1\over e^{\beta E_\pm(1- v/3)+1}} \ee
for the $(\pm)$ modes respectively. In the broken phase the corresponding
quantities $n_\pm^b$ are given by the same expressions with the reversal of
sign of $v$.

\vspace{.8cm}
\section{Solution inside the bubble wall}
\smallskip
We solve equation (2) in a perturbation series in the mass function. Assume
\[\Phi(z)=e^{3i(E-E_0+i\gm)z\sg_3}\Psi(z),\; 0\leq z\leq z_0\]
$\Psi(z)$ will then satisfy \[ {d\Psi\over dz}=iR(z)\Psi.\]
$R(z)$ has only off-diagonal elements,
\[ R(z)=\lt({0\atop M(z)}\qquad {\td{M}(z)\atop 0}\rt),\]
where $M(z)=3m(z)e^{6i(E-E_0+i\gm)z},$ and the tilde stands for
complex conjugation and change in the sign of $\gm$. We now convert it into
an integral equation,
\[\Psi (z)=\Phi(0)+i\int^z_0R(z^{\pr})\Psi(z^{\pr})dz^{\pr} \]
It has an iterative solution, $\Psi(z)=\Sigma(z)\Phi(0)$, where
\[\Sigma(z)=1+i\int^z_0dz^\pr R(z^\pr)-\int^z_0dz^\pr\int^{z^\pr}_0dz^{\pr\pr}
R(z^\pr)R(z^{\pr\pr})+\cdots \]
We shall actually need the solution for $\Phi(z)$ at $z=z_0$,
\be \Phi(z_0)=e^{3i(E-E_0+i\gm)z_0\sg_3}\Sigma(z_0)\Phi(z_0)\equiv \Omega(z_0)
\Phi(z_0).\ee
Writing \[\Omega(z_0)=\lt({\al\atop\td{\beta}}\qquad
{\beta\atop\td{\al}}\rt),\] we get
\br  \al &=& F(1+\int^{z_0}_0dz^\pr\int^{z^\pr}_0dz^{\pr\pr}\td{M}(z^\pr)
M(z^{\pr\pr})+\cdots)  \\ \beta &=& iF(\int^{z_0}_0dz^\pr\td{M}
(z^\pr)+\cdots) \er
with $F=e^{-3\gm z_0}e^{3i(E-E_0)z_0}$. Note that in the absence of damping
($\gm =0$), the tilde operation reduces to complex conjugation.

As already mentioned, simple extensions of the Higgs sector of the standard
model can provide an additional source of large CP violation for baryogenesis
In the standard model with a single Higgs doublet, the expectation value of the
Higgs field is real everywhere during the phase transition. But in multi-Higgs
models, some of the components acquire space dependent values within the bubble
wall. This in turn leads to complex space dependent mass function for the
quarks having Yukawa couplings to those multiplets. These are, in principle,
calculable from the model considered but, in practice, will depend on the many
Higgs self-couplings.
Here we avoid this problem by assuming a simple but anticipated form the mass
function,\be m(z)={m_0\over z_0}z+i{\delta\over z_0^2}z(z_0-z) ,\ee
within the bubble wall. The parameter $\delta$ relates to the CP violation
in the model.

It suffices to work out the perturbation series to second order to get the
leading contribution to the asymmetry in the baryonic currents. With the
parametrization (16), $\al$ and $\beta$ in (14-15) can be obtained explicitly
to this
order as
\br \al &=& F(1+9z_0^2U+\cdots)\no \\ \beta &=& iF(3z_0V+\cdots) \er
$U$ is quadratic in $m_0$ and $\delta$ and $V$ is linear,
\br  U &=& Am_0^2+iBm_0\delta +C\delta^2\no \\ V &=& am_0-ib\delta .\er
Each of the coefficients $A,a$ $etc.$ depend only on
$\sg=6 z_0\{\gm-i(E-E_0)\}$,
\br  A&=&e^{\sg}\lt({1\over \sg^3}-{1\over \sg^4}\rt)-{1\over 3\sg}-{1\over
2\sg^2}+
{1\over \sg^4},\no \\
B&=& e^{\sg}\lt({1\over \sg^3}-{4\over \sg^4}+{4\over \sg^5}\rt)+{1\over
3\sg^2}+
{1\over \sg^3}-{4\over \sg^5},\no \\
C&=& e^{\sg}\lt({1\over \sg^4}-{4\over \sg^5}+{4\over \sg^6}\rt)-{1\over
30\sg}+
{1\over 3\sg^3}+{1\over \sg^4}-{4\over \sg^6}, \no \\
a&=& e^{\sg}\lt({1\over \sg}-{1\over \sg^2}\rt)+{1\over \sg^2}, \no \\
b&=& e^{\sg}\lt({1\over \sg^2}-{2\over \sg^3}\rt)+{1\over \sg^2}+
{2\over \sg^3}  \er
\vspace{.8cm}
\section{Damping in reflection and transmission}
\smallskip
The damping rate has its origin in the scattering of the fermion under
consideration by other particles in the plasma. Although it does not alter
the densities of particles in the unbroken or broken phase, a quantum
mechanical amplitude gets attenuated as it traverses the plasma. In the
present context only the attenuation within the bubble wall is relevant,
where the baryon asymmetry is produced. Thus we prepare our states near the
wall of the bubble \cite{fn3}.

To study damping effects, consider, for example, the quasiparticle propagation
in the
normal mode. We send a right-handed fermion towards the domain wall from the
unbroken phase. Noting the reversal of chirality after reflection at the wall,
the incident wave (of unit current at $z=0$) and a reflected wave of amplitude
$r$, say, is given by
\be \Phi(z)=\lt({1\atop 0}\rt)e^{iKz}+\lt({0\atop r}\rt)e^{-iKz},\; z\leq 0 \ee
On the right (broken phase), we have only the transmitted wave of amplitude
$t$, say. From (10) we get
\be \Phi(z)=t\lt({c\atop -s}\rt)e^{iP(z-z_0)}, z\geq z_0 \ee
Eq.(13) now serves as the matching condition needed to find the reflection and
transmission amplitudes
\[t\lt({c\atop -s}\rt)=\lt({\al\atop\td{\beta}}\qquad {\beta\atop\td{\al}}\rt)
\lt({1\atop r}\rt), \] giving \be r=-(s\al+c\td{\beta})/D,\; t=1/D,\;
D=c\td{\al}+s\beta.\ee

In the absence of damping ($\gm=0$), the tilde operation on $\al$ and $\beta$
reduces to complex conjugation and we immediately obtain from (22),
\be |r|^2+|t|^2=1 \qquad  (\gm=0), \ee
expressing the equality of currents in the two phases. However, in the general
case ($\gm\neq 0$), this does not hold. The reflection and transmission
coefficients can be found in the general case to the second order in the mass
variables using the results for $\al$ and $\beta$ obtained in (17-19).
The expression are rather clumsy and will not be presented
here. But for large damping, i.e. large values of ($6\gm z_0$), the leading
behaviour of
these coefficients is simple to state,
\be |r|^2\longrightarrow O\lt(1\over (6\gm z_0)^4\rt),\qquad
|t|^2\longrightarrow O\lt(e^{-6\gm z_0}\rt) \ee
badly violating the current conservation relation (23).

Thus the damping causes the calculational scheme to violate unitarity and
results derived from such a scheme may be questioned. A satisfactory scheme
can only be found in a many body formulation of the problem which incorporates
the true mechanism responsible for damping.
\vspace{.8cm}
\section{Problem of baryogenesis}
\smallskip
Unitarity can be restored in the present framework only if the thickness of
the bubble wall tends to zero (thin wall approximation). In this limit the
quasiparticles would not have to cross any intervenning medium to go from one
phase to the other, which is precisely the region where the loss in probability
due to the damping matters in the present problem.

As $z_0\rightarrow 0$, the functions $A,B,C,a$ and $b$ approach constants and
we get
\br \al &=&1+9z_0^2({1\over 6}m_0^2+{i\over 30}m_0\delta+{1\over 72}
{\delta}^2) + O(z_0^3) \\ \beta&=& 3z_0({1\over 6}\delta+{i\over 2}m_0)
+O(z_0^2) \er
Accordingly the matrix $\Omega(z_0)\rightarrow 1$ and no asymmetry between
fermion and antifermion currents would result.

Nevertheless, we plan to calculate the baryon asymmetry in this model
for  zero wall thickness to examine the effect of damping it. For
this purpose we imagine that $\delta$ is very large so that $\delta z_0$ goes
to a non-zero constant, say,
$\Delta_0<1$, even if $z_0$ is very small. Then
\br \al&=& 1+{\Delta_0^2\over 8}+O(\Delta_0^3) \\ \beta&=&{1\over 2}
\Delta_0+O(\Delta_0^2) \er
Note that they are real and independent of $\gm$.

Since baryon non-conservation through sphaleron processes involves the
left-handed fermions and antifermions, we are interested in calculating only
the left-handed baryonic currents in the unbroken phase.

Consider the propagation of quark excitation in the normal mode. We have
already calculated for finite wall thickness the reflection and transmission
amplitude when right-handed fermions are incident on the wall from the
unbroken phase. We rewrite them for thin wall approximation,
\[r=-(s\al+c\beta)/D,\qquad  t={1\over D},\qquad  D=c\al+s\beta.\]
With (25) and (26) the reflection coefficients become,
\[ T_+=|t|^2=1/|D|^2,\qquad R_+=1-|t|^2.\]
where \[|D|^2=1+hm_0^2+{\Delta_0^2\over 4}+{2\over 3}hpm_0\Delta_0,\qquad
h^{-1}=4\lt(\gm^2+(p/3)^2\rt)\]
The incident flux is the same for particles and antiparticles, {\it viz.}
${1\over 3}n^u_+$, where $n^u_+$ is given by (12).
Considering both the particles and antiparticles, the net contribution to the
reflected left handed baryonic current is \cite{fn4}
\be \int {dk\over 2\pi}{1\over 3}n^u_+(R_+-\bar{R}_+)\ee

Here and in the following a bar on a reflection or transmission coefficient
denotes the corresponding quantity for the antiparticle. It is obtained by
solving the same eqn.(2) with $m$ replaced by $m^{\ast}$.

Next consider transmission in the unbroken phase due to incidence on the wall
from the brooken phase. On the left there is only a transmitted wave of
amplitude $t^\pr$, say \[ \Phi (z)=\lt({0\atop {t^\pr}}\rt)e^{-iKz},\; z<0\]
On the right we have both the incident wave and the reflected wave of
amplitude $r^\pr$, say.
\[ \Phi (z)=\lt(s\atop -c\rt)e^{-iPz}+r^\pr \lt(c\atop -s\rt)e^{iPz},\;z>0 \]
Again the matching condition (13) gives \[\lt({s+r^\pr c\atop {-(c+r^\pr s)}}
\rt)=\lt({\al \atop\beta} \qquad {\beta\atop\al}\rt) \lt(0\atop t^\pr\rt) \]
giving \[r^\pr =-(s\al+c\beta)/D,\qquad t^\pr =(s^2-c^2)/D \]
Note that the coefficients fail to satisfy the current conservation across
the wall, $|r^\pr|^2+|t^\pr|^2\neq 1$. The problem here is that the current in
the broken phase as calculated from the wave function using (4) is
not $-1+|r^\pr|^2$ but
contains cross terms,
\bq j&=&-1+|r^\pr|^2+\{r^\pr (s^{\ast}c-sc^{\ast})+c.c.\}\no \\
&=&-|1-r^\pr (s^{\ast}c-sc^{\ast})|^2+|r^\pr|^2\{1+|s^{\ast}c-sc^{\ast}|^2\}
\no \\ &=&-|c^2-s^2|^2+|r^\pr|^2|c^2-s^2|^2 ,\eq
on using $|c|^2-|s|^2=1$. Thus with correct normalization, these
reflection and transmission coefficients coincide, as usual, with those
calculated for incidence from the unbroken phase. The transmitted left handed
baryonic current in the unbroken phase is given by
\be \int {dp\over 2\pi} {\pl E_+\over \pl p} n_+^b (T_+-\bar{T}_+) \ee
where $\pl E_+/\pl p$ is the group velocity in the broken phase.
Adding (29) and (30) we get the total baryonic current in the normal mode in
the
unbroken phase due to reflection and transmission as
\be J_+= \int {dk\over 2\pi}{1\over 3} (n_+^b-n_+^u) (T_+-\bar{T}_+) \ee

In a similar way we can work out the baryonic current in the abnormal mode in
unbroken phase to be
\be J_-= \int {dk\over 2\pi}{1\over 3} (n_-^b-n_-^u) (T_--\bar{T}_-) \ee
where \[T_-=1/{|\al c^\pr - \beta s^\pr|^2}\] and $c^\pr$, $s^\pr$ are
given by same expressions as for $c$ and $s$ with changes of sign of
$(E-E_0)$ and of $\phi$.

It is now easy to make an order of magnitude estimate of these currents.
Formally the $k$- integrals in (31) and (32) extend over $0$ to $\infty$ and
$0$
to $3E_0$ respectively. But the integrands are highly damped at higher
values of $k$, not just becuse of the presence of the density functions; the
transmission coefficients for a realistic (i.e. smooth and finite width)
bubble wall would have fallen exponentially for even lower $k$ values. We set
$E_0$ as a resonable upper limit for both the integrals. Hopefully the low
momentum approximation, on which the effective Lagrangian (1) is based, would
admit this upper limit.

As the temperature of the phase transition is about $100$ GeV, we may
expand the density functions in $v$ for small $v$ and approximate the
exponential by unity to get
\be n_{\pm}^b-n_{\pm}^u\simeq -{1\over 6}\beta v E_\pm \ee
Thus the currents (31) and (32) become
\be J_\pm =\pm {\beta vm_0\Delta_0\over 2+\Delta_0^2}
\int^{E_0}_0 {dk\over 2\pi}\,(E_0\pm {k\over 3})\,{p\over p^2+9\gm^2} \ee

Observe the large cancellation in the sum of $J_+$ and $J_-$ giving the total
CP-violating left-handed baryonic current in the unbroken phase.
Evaluating the resulting integral approximately we get
\br J^L_{CP}=J_++J_- \no \\ \simeq {\beta vm_0\Delta_0\over 2+\Delta_0^2} \er

The final step is to obtain the baryonic density $n_B$ in the broken phase
from the steady state solution to the rate equations in the two phases.
CKN\cite{ckn} find numerical solution to the Boltzmann equation. We shall
follow FS \cite{fs}, who solve the diffusion equations for small bubble
wall velocity to get \be n_B=J^L_{CP}f \ee
where $f$ is a given function of the diffusion coefficients for quarks and
leptons, the wall velocity and the sphaleron induced baryon number violation
rate. Their estimate for $f$ is $10^{-3}\leq f\leq 1$ in MSM, which should
also be valid for its simple extensions.

Noting the one dimensional entropy density $s=73\pi/3\beta$, the baryon to
entropy ratio is obtained as
\be n_B/s\sim 1.3\times 10^{-2}v\beta^2 m_0\Delta_0 f/{(2+\Delta_0^2)} \ee

With $m_0=5\, GeV, \beta=10^{-2} GeV^{-1}, v=0.1$ and $\Delta_0=1$, we get
$n_B/s\sim 3\times 10^{-7}f$, to be compared with the observed value
$n_B/s \sim 5\times 10^{-11}$.

\vspace{.8cm}
\section{Conclusion}
\smallskip
We have studied the effect of damping on quasiparticle propagation in the
electroweak plasma in connection with the problem of baryogenesis in non-
minimal versions of the standard model, where the Higgs sector is extended
to include more than one multiplet. In general, such an extension gives
rise to a complex mass function for a quark within the bubble wall formed
during the phase transition. This constitutes the CP violation needed
for baryogenesis. Its real and imaginary parts are parametrized in a simple
way making the integrals simple to evaluate. We follow closely the technique
of CKN\cite{ckn}, but consider a direct separation of baryon number by the
bubble wall rather than of some other CP-odd charge \cite{shp}.

The inclusion of the temperature dependent effective mass gives rise to two
modes of quasiparticle propagation in the plasma. Our calculation shows that
both modes must be taken into account. In fact, the net baryon asymmetry
current results after large cancellation between the baryonic currents
carried separately by the two modes.

We also include the recently discussed damping suffered by the quasiparticles
while propagating through the plasma \cite{ghop,hs}. Like the effective mass,
it also arises
from interaction of a fermion with other particles in the plasma. This damping
appears to ruin an otherwise successful calculation \cite{fs} of
baryon asymmetry in the
MSM. Besides spatial attenuation, it also affects the spinor components of
$\Phi$ and the dispersion relations are no longer separated by a mass gap.
Also the two  spinor components acquire equal and opposite phase.

The quasiparticle propagation is damped everywhere within the medium, in the
broken and unbroken phases, as well as within the bubble wall. However, it is
the damping associated with the propagation within the wall, which is of
concern in the problem of baryogenesis. It causes the magnitudes of the
reflection coefficients to be reduced, more so the larger the value of
$\gm z_0$. The problem one faces here is that the calculational
scheme is non-unitary and the question of reliability of a calculation
within such a scheme remains open.

Unitarity is restored in this formalism only in the thin wall approximation,
when a calculation of baryon asymmetry would be free from the above objection.
But $\Omega(z_0)\rightarrow 1$ as $z_0\rightarrow 0$ and there is no asymmetry
between the fermion and the antifermion as they interact with the wall. We
are then led to consider a rather unrealistic situation where the coefficient
$\delta$ in the imaginary part of the mass function is so large that
$\delta z_0=\Delta_0$, a non-zero constant, even in the thin wall
approximation. It is now simple to calculate the baryon to entropy ratio of
the Universe. As expected, it does not suffer any reduction due to damping
rate when compared with a similar expression calculated without taking it into
account \cite{sm}

We now come back to comment on the real problem with finite wall thickness.
Though a satisfactory framework for calculation should avoid any non-unitarity
due to damping by including the many body processes responsible for it, it,
nevertheless, appears that the damping effect in baryon asymmetry will
persist even in such a framework. Thus the result calculated without
damping \cite{sm} is likely to be reduced by suppression factor appearing in
the
reflection and transmission coefficients (24).
Note that this suppression is much less than that
encountered in MSM where one has to go to higher order in perturbation
expansion in the mass matrix. Furthermore, the magnitude of the wall
thickness in extended versions of the standard model is not known.
In view of the many parameters in the potential of such a model, the
indication is that it could be small in several regions of the parameters.
For $z_0\leq {1\over 15}GeV^{-1}$, the damping factor is of order $10^{-4}$
at most. Within the uncertainties of the transport properties of the
electroweak plasma, the baryon to entropy ratio would still be around the
observed magnitude.

CKN \cite{ckn} find that the t quark propagation through the bubble wall can
produce
the observed baryon to entropy ratio. We here argue that the contribution
of the b quark, for which finite temperature correction in mass, in both the
real and imaginary parts, is important, is also likely to be of similar
magnitude.

\end{document}